\documentclass[10pt,twoside]{article}
\usepackage{epsf,colordvi,graphicx,color,amsbsy}
\begin{document}
\title{\bfseries A universal law for tails of density pdf's in
  multi-dimensional Burgers turbulence}
%
\author{ J.~Bec$^{1,2}$, U.~Frisch$^{1,2}$, K.~Khanin$^{3}$ and
  B.~Villone$^{4}$}
\date{}
\maketitle
\vspace*{-9 mm}
{\small
\begin{center}
$^1$Observatoire de la C{\^o}te d'Azur, Lab. G.D. Cassini\\
B.P. 4229, F-06304 Nice Cedex 4, FRANCE
\\
$^2$CNLS - Theoretical Division, LANL, Los Alamos, NM~87545, USA
\\
$^3$Isaac Newton Institute for Mathematical Sciences\\
20 Clarkson Road, Cambridge CB3 0EH, UK
\\
$^4$CNR - Istituto di Cosmogeofisica, 10133 Torino, ITALY
\end{center}
}
\noindent
Contact e-mail: {\texttt bec@obs-nice.fr}
\thispagestyle{empty}
\begin{abstract}
  \noindent Extending work of E, Khanin, Mazel and Sinai \cite{ekms97}
  on the one-dimensional Burgers equation, we show that density pdf's
  have universal power-law tails with exponent -7/2. This behavior
  stems from singularities, other than shocks, whose nature is quite
  different in one and several dimensions. We briefly discuss the
  possibility of detecting singularities of Navier--Stokes turbulence
  using pdf tails.
\end{abstract}

\section{Introduction}
In recent years there has been considerable interest in probability
density functions (pdf) for Navier--Stokes turbulence. Similar
questions can be asked for random solutions of Burgers equation
(``burgulence'').  We are interested here in the tail behavior of the
pdf of the density $\rho$ for solutions to the $d$-dimensional
Burgers equation in the limit of vanishing viscosity ($\nu\to 0$):
\begin{eqnarray}
&&\partial_t {\bf v} + ({\bf v}\cdot\nabla){\bf v} = \nu \nabla^2 {\bf v},
\quad {\bf v} = - \nabla \psi,
\label{burgdd}\\
&&\partial_t \rho + \nabla\cdot\left(\rho {\bf v}\right)= 0.
\label{rhodd}
\end{eqnarray}
The initial potential $\psi_0({\bf r}_0)$ and the initial density
$\rho_0({\bf r}_0)$ are random functions of the space variable. This
problem arises, for example, in the study of large-scale structures
in the Universe (see Ref.~\cite{3} and references therein).  As is
well known, the Burgers equation leads to shocks in which the density
of matter is infinite. Yet, large but finite densities do not
necessarily occur in the neighborhood of shocks. E {\it et al.}
\cite{ekms97} considered a related problem of determining the pdf of
the velocity gradient for the one-dimensional Burgers equation with
large-scale and white-in-time random forcing. They showed that large
negative gradients come from ``preshocks'' (nascent shocks) which
contribute a power-law tail with exponent -7/2 to the pdf. Preshocks
correspond to fast fluid particles catching up for the first time with
slow ones. They constitute discrete events in (Eulerian) space time.
We have shown that the -7/2 law for the pdf of (negative) velocity
gradients in one dimension applies also (i) for decaying (unforced)
``burgulence'' with smooth (i.e. large-scale) random initial conditions
\cite{1} and (ii) for the case of a deterministic time and
space-periodic force which is a sum of delta functions in time with
smooth space dependence \cite{2}.  For this ``kicked burgulence'', the
-7/2 law was also obtained numerically (see Fig.~1).
\begin{figure}[t]
  \begin{center}
    \includegraphics[width= 0.6 \linewidth]{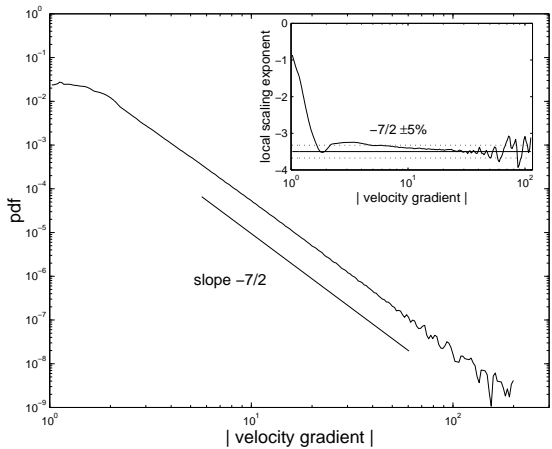}
  \end{center}
  \label{f:first-deriv}
  \caption{Pdf of the velocity gradient at negative values in log-log
    coordinates, for one-dimensional kicked burgulence.  A power law
    with exponent $-7/2$ is obtained over two decades. Simulation
    uses a modification of the Fast Legendre
Transform algorithm of Ref.~\cite{nv94}.}
\end{figure}

\section{``Kurtoparabolic'' points and the -7/2 law}
We turn now to the multi-dimensional problem
(\ref{burgdd})-(\ref{rhodd})
with smooth and random initial conditions (for details see Ref.~\cite{3}).
As is known, the one or multi-dimensional viscous Burgers equation
can be solved explicitly by means of the Cole--Hopf transformation.
From this, by taking the limit $\nu\to 0$, a ``maximum
representation'' can be derived for the velocity potential
\begin{equation}
\psi({\bf r}, t) =\max_{{\bf r}_0} \left (\psi_0({\bf r}_0)
- \frac{|{\bf r} - {\bf r}_0|^2}{2t} \right).
\label{maxrep}
\end{equation}
The maximum is achieved at a point at which the gradient of the r.h.s.
vanishes, leading to ${\bf r}={\bf r}_0+t{\bf v}({\bf r}_0,0)$. From
this it follows that ${\bf r}_0$ is a Lagrangian coordinate. The
``naive'' Lagrangian map ${\bf r}_0\mapsto {\bf r}$ given by this
relation is however not invertible, except for short times.  The
requirement that the maximum in (\ref{maxrep}) is {\em global\/} can
be recast in geometrical terms by introducing the Lagrangian potential
$\varphi({\bf r}_0, t) \equiv t\psi_0({\bf r}_0) - |{\bf r}_0|^2/2$ and
the ``proper'' Lagrangian map ${\bf r}_0\mapsto {\bf r} \equiv-\nabla
\varphi_c({\bf r}_0, t)$, where $\varphi_c$ is the convex-hull of
$\varphi$ with respect to ${\bf r}_0$. For example, in one dimension,
the graph of $\varphi_c$ is obtained by tightly pulling a string over
the graph of $\varphi$. The graph of $\varphi_c$ coincides with that
of $\varphi$ at regular points, wherever fluid particles have not yet
fallen into shocks. It also contains linear and ruled manifolds
associated to the different types of shocks: segments when $d=1$,
triangles and ruled surfaces when $d=2$, etc. Conservation of mass
implies that the density is given, at regular points, by $\rho({\bf
  r}, t) = \rho_0({\bf r}_0)/J({\bf r}_0,t)$, where $J$ is the
Jacobian of the Lagrangian map. (The density is infinite in shocks.)
Since the Jacobian is (up to a factor $(-1)^d$) equal to the Hessian
of the Lagrangian potential (determinant of the matrix of second space
derivatives), it follows that large densities are typically obtained
only near parabolic points. However, arbitrarily close to a parabolic
point there are generically hyperbolic points where the surface
defined by $\varphi$ crosses its tangent (hyper)plane and which,
therefore, do not belong to its convex hull.  Yet, there exist in
general exceptional ``kurtoparabolic'' points which are parabolic and
belong to the boundary of the set of regular points. Near such points,
arbitrarily large densities are obtained.  In one dimension, the only
kurtoparabolic points are the preshocks which are discrete space-time
events in both Eulerian and Lagrangian coordinates. In two and more
dimensions, kurtoparabolic points are also born at preshocks but
persist in general for a finite time (see Fig.~2). In Eulerian space,
they are associated to boundaries of shocks (e.g. end points of shock
lines for $d=2$).
\begin{figure}[h!]
\begin{minipage}[c]{.45 \linewidth}
\begin{center}
\includegraphics[width=\linewidth]{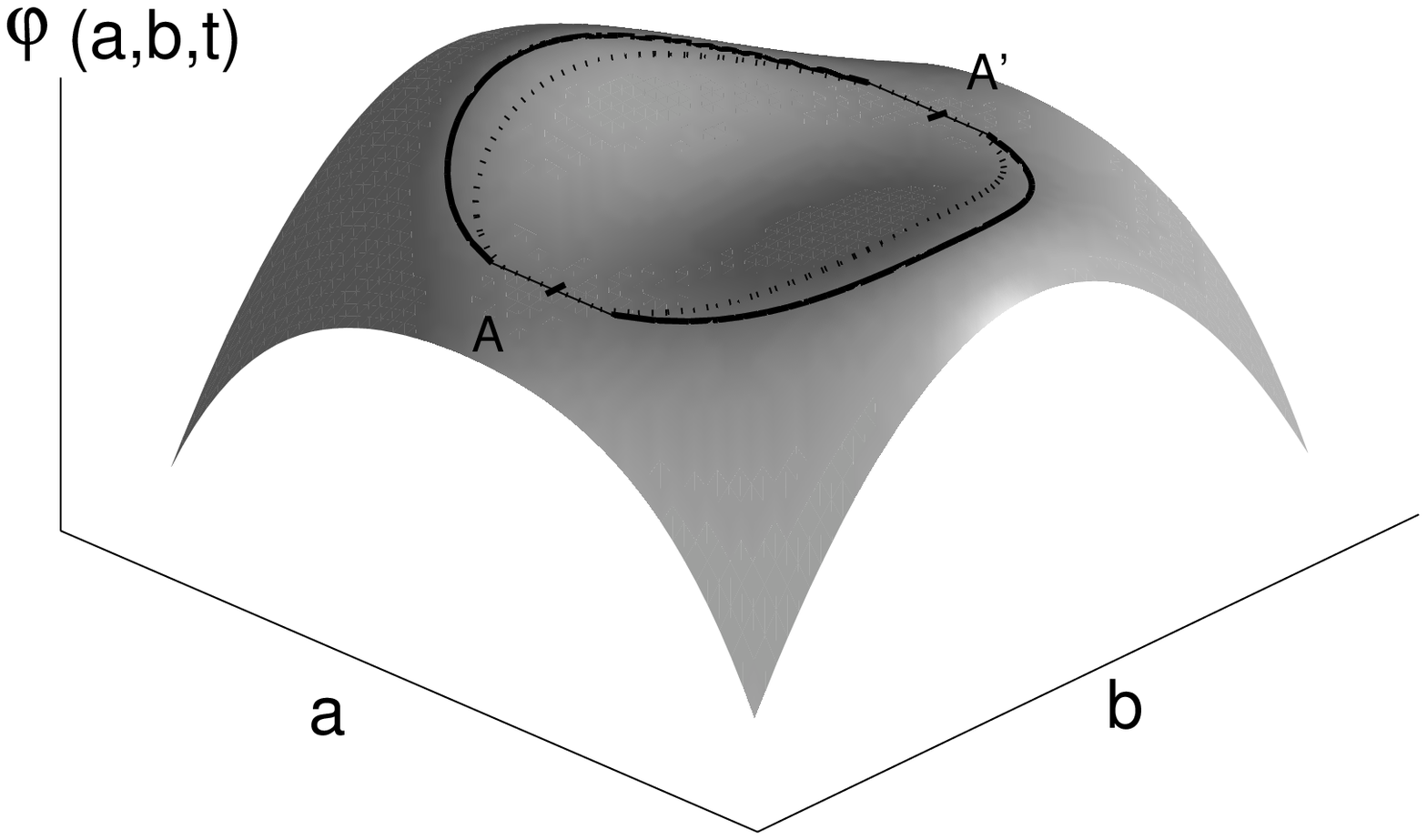}
\end{center}
\end{minipage}
\hfill
\begin{minipage}[c]{.45 \linewidth}
\begin{center}
\includegraphics[width=\linewidth]{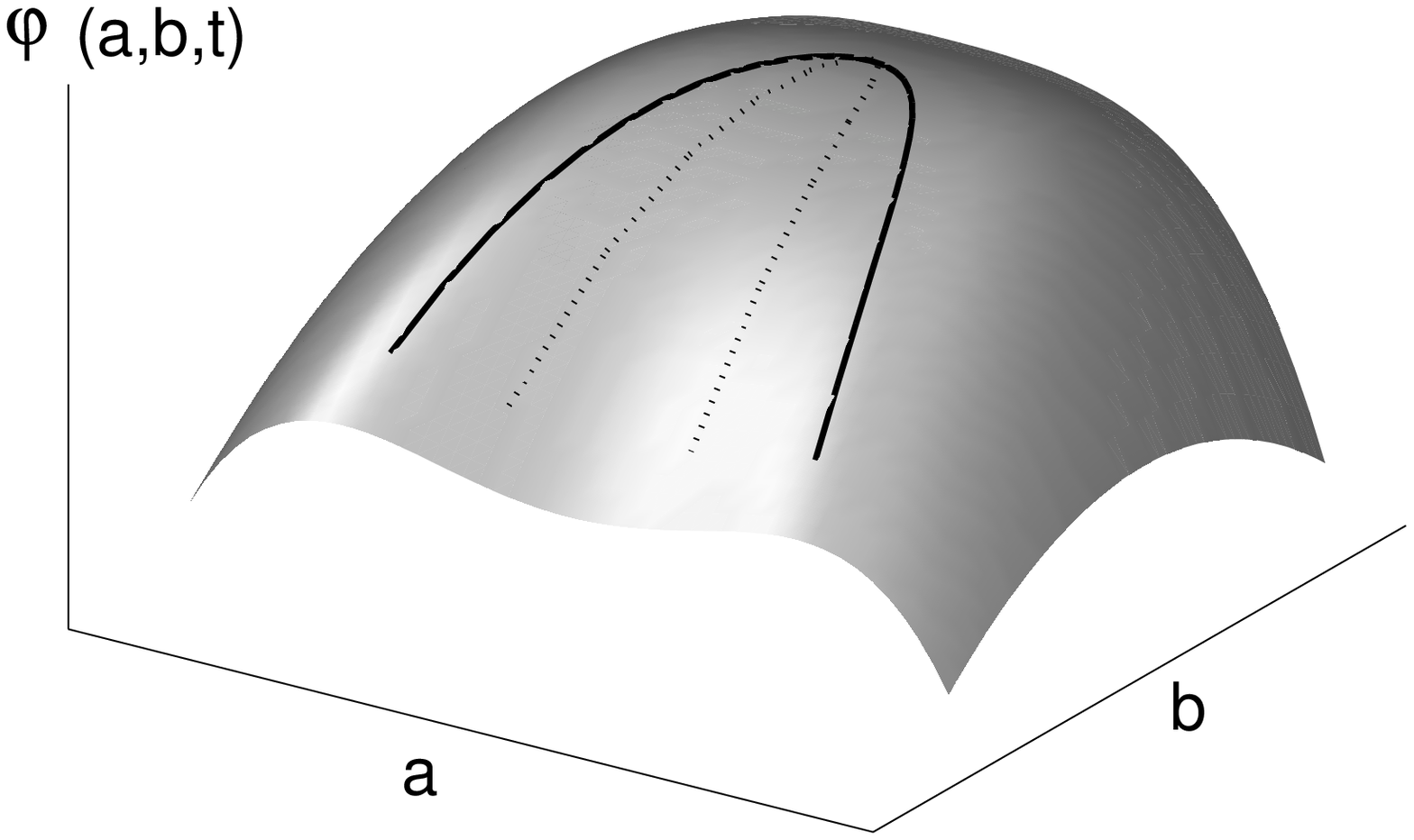}
\end{center}
\end{minipage}
\caption{The Lagrangian potential in two dimensions, just after a 
preshock (left) and in the neighborhood of a kurtoparabolic
point (right). Continuous lines: separatrices between
the regular part and the ruled surface of the convex hull; 
dotted lines: vanishing of the Jacobian of the
Lagrangian map. A and A' are a pair of  kurtoparabolic points
born with the shock.}
\end{figure}

We just indicate general ideas involved in the determination of the
density pdf for random initial conditions (which need not be
homogeneous). The determination of the large-$\rho$ tail of the
cumulative probability distribution of the density is equivalent to
finding the fraction of Eulerian space-time where $\rho$ exceeds a
given value.  The latter is determined by changing from Eulerian to
Lagrangian coordinates and using a suitable normal form (i.e.\ a Taylor
expansion to the relevant order) of the
Lagrangian potential near a kurtoparabolic point.  The theory is
rather different in one dimension and higher dimensions, because
kurtoparabolic points are persistent only in the latter case.
However, the scaling law for the resulting pdf, namely $\propto
\rho^{-7/2}$, is the same in all dimensions. In fact, when
$d\ge 2$, two orthogonal spatial directions play the same role as
space and time in one dimension.
  
\section{Detecting Navier--Stokes singularities}
It is clear that the algebraic tail of the pdf of velocity gradients
or of density for burgulence comes from identified singularities.
Measurements of pdf's for space or time derivatives of Eulerian
velocities for incompressible three-dimensional Navier--Stokes
turbulence have not revealed  power law tails, but such tails may just
have been, so far, ``lost in the
experimental noise''. There has indeed been considerable speculations 
about singularities of the
Navier--Stokes equations in the inviscid limit. If singularities with
divergent gradients are present, they will give  power-law tails, 
at least as intermediate asymptotics when the viscosity is small (the
converse is however not true, since statistical effects not related
to singularities can also give power laws). The confirmed absence of power laws
would probably rule out singularities.

\end{document}